\documentclass[runningheads]{llncs}
\usepackage{graphicx}

\usepackage{tikz}
\usepackage{comment}
\usepackage{amsmath,amssymb} 
\usepackage{color}
\usepackage{adjustbox}
\usepackage[accsupp]{axessibility}  
\usepackage{cite}

\begin{document}
\pagestyle{headings}
\mainmatter
\def\ECCVSubNumber{100}  

\title{Adaptive GLCM sampling for transformer-based COVID-19 detection on CT } 


\titlerunning{Adaptive GLCM sampling for transformer-based COVID-19 detection on CT}
%
\author{Okchul Jung\inst{1}\thanks{Authors contributed equally.} \and
Dong Un Kang\inst{1}$^\star$ \and
Gwanghyun Kim\inst{1}$^\star$ \and
Se Young Chun\inst{1,2}}

\authorrunning{Jung et al.}
%
\institute{Electrical and Computer Engineering, Seoul National University, Korea\\
\and
 Interdisciplinary Program in AI, Seoul National University, Korea\\
\email{\{luckyjung96, qkrtnskfk23, gwang.kim, sychun\}@snu.ac.kr}}

\maketitle

\begin{abstract}
The world has suffered from COVID-19 (SARS-CoV-2) for the last two years, causing much damage and change in people’s daily lives. Thus, automated detection of COVID-19 utilizing deep learning on chest computed tomography (CT) scans became promising, which helps correct diagnosis efficiently. Recently, transformer-based COVID-19 detection method on CT is proposed to utilize 3D information in CT volume. However, its sampling method for selecting slices is not optimal. To leverage rich 3D information in CT volume, we propose a transformer-based COVID-19 detection using a novel data curation and adaptive sampling method using gray level co-occurrence matrices (GLCM). To train the model which consists of CNN layer, followed by transformer architecture, we first executed data curation based on lung segmentation and utilized the entropy of GLCM value of every slice in CT volumes to select important slices for the prediction. The experimental results show that the proposed method improve the detection performance with large margin without much difficult modification to the model.

\end{abstract}

\section{Introduction}

With the development of deep learning, it has been applied to many research areas including medical image analysis. Especially, COVID-19 (SARS-CoV-2) is a respiratory disease, which has caused much damage globally due to its contagious characteristic. Since it is a respiratory disease, it is hard to differentiate from other respiratory diseases because they have similar symptoms \cite{ng2020covid}.
Therefore, computed tomography (CT) scans become promising for COVID-19 detection thanks to its high sensitivity and specificity \cite{bernheim2020chest}. With the help of deep learning, it is expected to diagnose patients with COVID-19 on CT scans correctly and efficiently. CT-scans are often treated as three-dimensional data. There have been attempts that try to utilize Convolutional Neural Network in trying to capture three-dimensional data by having 3D convolution \cite{konar2021auto, mohammed2020weakly}. However, the computation complexity of 3D convolution is higher than conventional 2D convolution, limiting its use. Although utilizing 2D information of CT scans seems most likely, the symptoms of COVID-19 might underlie in the 3D information \cite{kwee2020chest}. 

Fortunately, COV19-CT-DB database  which is annotated for COVID-19 detection, consisting of about 7,700 3-D CT scans was given by the hosts of the ECCV COVID-19 detection challenge \cite{kollias2018deep, kollias2020transparent, kollias2020deep, kollias2021mia, kollias2022ai}. The dataset includes 3D information providing semantic features to utilize in the diagnosis of COVID-19. Treating the dataset at slice-level would not give a significant result and not be able to fully take advantage of the semantic information within the dataset. Recently, COVID-19 detection method using transformer architecture \cite{vaswani2017attention} as well as convolutional neural network (CNN) is proposed to use 3D information in CT volume efficiently \cite{hsu2021visual}. However, the use of the 3D context of CT volumes in its work is limited by including uninformative slices which don't include lungs, and selecting the only slices near the center points. 

To fully use 3D information in CT volumes, we propose a transformer-based COVID-19 detection on CT volumes with novel data curation and adaptive sampling method using gray level co-occurrence matrices (GLCM). We mainly focused on sampling adequate data to exclude unnecessary and redundant data that is not helpful for the model to learn the correct COVID-19 prediction. To achieve this, based on the combination of CNN and transformer in \cite{hsu2021visual}, we propose a novel data curation through lung segmentation. Furthermore, we calculate the entropy of GLCM to choose more significant slices for the prediction during the sampling. Our extensive experiments confirm that the proposed method significantly improves the COVID-19 detection performance on the validation set.

\begin{figure}[!t]
    \centering
    \includegraphics[width=0.9\linewidth]{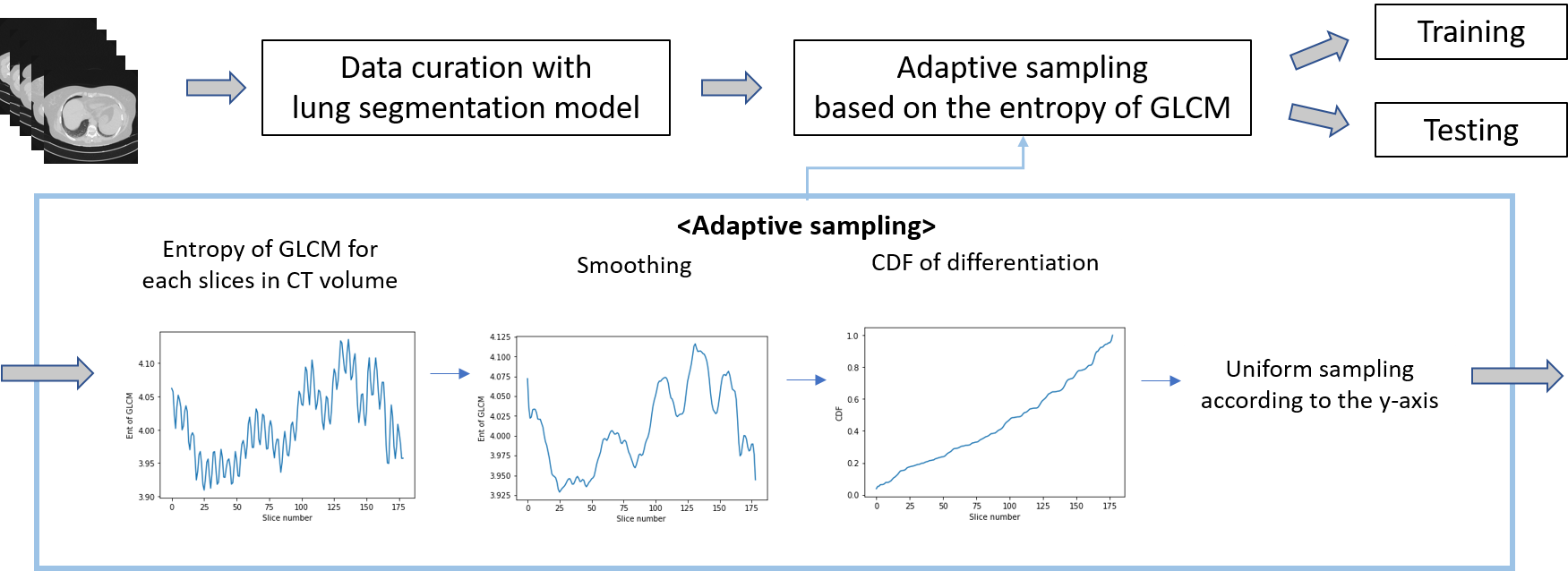}
    \caption{Flowchart of the proposed data curation with the pretrained lung segmentation model and adaptive sampling based on the entropy of GLCM.}
    \label{fig:method}
\end{figure}

\section{Proposed Method}

Our model framework was initially inspired by the framework suggested by participants in the 1st year competition. We adopt CT-scan aware transformer (CCAT) architecture, a combination of CNN and transformer as proposed in \cite{hsu2021visual}. In specific, the entire model is consisted of CNN and a transformer to fully utilize the semantic information within 3D CT scans. CNN can efficiently learn the feature representation of the images. Then, the within-slice transformer (WST), which learns the relation within pixels in each slice, is utilized followed by the between-slice transformer(BST), which aggregates the resulting features from each slice \cite{hsu2021visual}. By doing so, the entire model can see and learn the whole CT scans in three dimensions. Given COV19-CT-DB datasets are consisted of 3D CT scan volumes, not all the images were useful for the model to learn the COVID-19 prediction. Some images were given as a mixed combination of the horizontal and lateral views of the patients. Some images did not contain enough size of the lung, which makes it harder for the model to learn the COVID-19 prediction task. To mitigate this issue and fully leverage the 3D context of CT volumes, we propose a data curation with lung segmentation and adaptive sampling based on the entropy of GLCM as illustrated in Fig.~\ref{fig:method}.

Specifically, to selectively choose the most informative data for the model to learn, we decided to adequately pick the most useful data, in other words, an image that has a sufficient size lung. COVID-19 being a respiratory disease, selecting an image with sufficient size of the lung was conducted through lung segmentation. We heuristically curated data with lung segmentation on different percentages of how much lung occupies in the image. 

Furthermore, we propose an adaptive sampling based on the entropy of GLCM.
GLCM by definition is a matrix showing the different combinations of gray levels found within the image. Thus, it is able to describe the characteristics of an image by calculating how often pairs of the pixel with certain values and their spatial relationships occur in an image. This leads to the assumption that dissimilar sequential images are likely to have a steep difference in GLCM values. Our ultimate goal was to find the set of dissimilar images in a CT volume because dissimilar images are likely to be informative for the model to learn the covid detection task, making it robust. Therefore, we needed to pick sequential images that showed steep GLCM value differences. In specific, we first calculate the entropy of GLCM per each CT slice. Raw entropy data includes too high variation in values between slides in CT scan. To alleviate this issue, we employed Savitzky-Golay filter~\cite{savitzky1964smoothing} to smooth the graph of entropy of GLCM overall CT slices under certain window sizes. The filter refers to neighboring values to average its values, enabling smoothing along with the slices. Then, we conduct the CDF generation process in the order of differentiation of the GLCM graph, calculation of absolute value and normalization, and then generation of CDF. Finally, we perform uniform sampling according to the CDF value between 0 to 1. The model is trained and evaluated With the proposed sampling methods on the curated dataset.

\section{Experiments}

\subsection{Experimental Settings}

As our model, we use CCAT with ResNet-50 \cite{he2016deep} backbone, followed by WST and BST \cite{hsu2021visual} which are 1-layer and 2-layer standard transformers with 1 head respectively. For training of the model, we use Adam optimizer with a learning rate of 1e-6 and batch size of 8. We trained the model until converged with the step learning rate scheduler. We first resize each slice in CT volume into the size of $256\times256$ and randomly cropped to generate $224\times224$ images. We perform random shift/scaling/rotation and brightness/contrast data augmentation. During the test time, we first resize each slice in CT volume to the size of $256\times256$ and cropped $224\times224$ images at the center of them. We utilize the macro F1 score of COVID-19 prediction as our main metric and additionally report sensitivity and specificity.

\begin{table}[!t]
\caption{Comparison of validation performance among the COVID-19 detection methods. The results are in \%. Center focused sampling is the method used in original CCAT work \cite{hsu2021visual}. $s$ represents the window size used in Savitzky-Golay filter.}\label{tab:comparision}
\centering
\begin{adjustbox}{width=0.62\linewidth}
\begin{tabular}{cccc}
\hline
Sampling method                  & Macro F1       & Sensitivity    & Specificity    \\ \hline
Center focused (no data curation) & 84.57          & 83.64          & 67.46          \\
Center focused                     & 89.05          & 81.13          & \textbf{96.28} \\
GLCM ($s=17$)             & 87.60          & 87.74          & 88.10          \\
GLCM ($s=9$)              & 87.80          & 87.26          & 89.22          \\
GLCM ($s=3$)              & \textbf{89.49} & \textbf{89.15} & 89.59          \\ \hline
\end{tabular}
\end{adjustbox}
\end{table}

\subsection{Data Curation}

Selecting the most informative and essential data for training the prediction model is considered important in our proposed model. Among the images in each CT scan, we picked an image depending on how much lung area occupies in the whole image and calculated its percentage. We heuristically chose the percentage through multiple experiments ranging from percentages of 0.5, 1, 2, 3, 4, and 5. Capturing the lung area was executed through lung segmentation, and the optimal percentage was derived as 5. If the percentage was too high, segmentation failed to capture the whole dataset, excluding too much of a training and validation dataset, worsening the generalization of the model. Conversely, if the percentage was too low, segmentation failed to exclude images that might hinder the model's learning of useful feature representations.

\subsection{Adaptive Sampling based on the Entropy of GLCM}

Utilizing adaptive GLCM sampling to train and test the data resulted in unbiased and balanced sensitivity and specificity overall. That is, the model can successfully capture the COVID-19 dataset and non-COVID-19 dataset without being biased to a certain dataset, making it robust to COVID-19 prediction under imbalanced dataset conditions. Conventional methods without GLCM tended to show high specificity and low sensitivity, leading to a low macro f1 score. In other words, conventional methods that do not use GLCM-based sampling are able to correctly identify true negative predictions most of the time as shown below in Table~\ref{tab:comparision}. However, the method does not show high performance when it comes to sensitivity, which is a measure of the ability to correctly identify true positive predictions. Moreover, we employed a Savitzky-Golay filter with varying window sizes of 3, 9, and 17 to smooth the graph of entropy of GLCM overall CT slices. We found that when the window size is 3, the model shows the optimal performance outperforming the previous center-focused sampling method \cite{hsu2021visual}.

\section{Conclusions}

In this paper, our model consists of CNN and a transformer with the intention to fully capture the semantic information within the 3D CT scans. Moreover, we proposed a novel data sampling strategy so that the model can only learn from helpful and meaningful data. The data sampling strategy consisted of data curation with lung segmentation and GLCM value-based sampling. The introduced data sampling can be easily applied and implemented on various medical imaging tasks. Also, it is expected to give unbiased, balanced results on both sensitivity and specificity, showing the promising result.

\clearpage
%

\bibliographystyle{splncs04}

\bibliography{egbib}

\end{document}